\documentclass[twocolumn,letter]{jpsj2}

\def\runauthor{M. Matsumoto and M. Koga}

\title{Exciton-Mediated Triplet Superconductivity in $T_h$ System PrOs$_4$Sb$_{12}$}

\author{Masashige Matsumoto and Mikito Koga$^1$}
\inst{Department of Physics, Faculty of Science, Shizuoka University, Shizuoka 422-8529 \\
$^1$Department of Physics, Faculty of Education, Shizuoka University, Shizuoka 422-8529}

\recdate{February 24, 2005}

\newcommand{\Pra}{PrOs$_4$Sb$_{12}$}

\abst{
In \Pra, the lowest-lying singlet and triplet states in a Pr $4f^2$ configuration
hybridize with conduction electrons having local $a_u$ and $t_u$ point-group symmetries.
It is shown that for an attractive triplet pairing interaction,
the orbital degrees of freedom of the $t_u$ component are important.
In addition, the $T_h$ point-group symmetry characteristic of skutterudites
plays an important role in stabilizing triplet superconductivity.
}

\kword{
skutterudite,
triplet superconductivity,
time reversal symmetry,
exciton
}

\begin{document}
\sloppy
\maketitle


\renewcommand{\theequation}{\arabic{equation}} 
\newcommand{\br}{{\mbox{\boldmath$r$}}}
\newcommand{\bk}{{\mbox{\boldmath$k$}}}
\newcommand{\sk}{{\mbox{\footnotesize $k$}}}
\newcommand{\bskp}{{\mbox{\footnotesize \boldmath$k$}}}
\newcommand{\bsrp}{{\mbox{\footnotesize \boldmath$r$}}}
\newcommand{\bsk}{\bskp}
\newcommand{\bsr}{\bsrp}
\newcommand{\bS}{{\mbox{\boldmath$S$}}}
\newcommand{\bs}{{\mbox{\boldmath$s$}}}
\newcommand{\bQ}{{\mbox{\boldmath$Q$}}}
\newcommand{\bq}{{\mbox{\boldmath$q$}}}
\newcommand{\bV}{{\mbox{\boldmath$V$}}}
\newcommand{\bt}{{\mbox{\boldmath$t$}}}
\newcommand{\bd}{{\mbox{\boldmath$d$}}}
\newcommand{\bsigma}{{\mbox{\boldmath$\sigma$}}}

\newcommand{\bhx}{\hat{\mbox{\boldmath$x$}}}
\newcommand{\bhy}{\hat{\mbox{\boldmath$y$}}}
\newcommand{\bhz}{\hat{\mbox{\boldmath$z$}}}
\newcommand{\hx}{\hat{x}}
\newcommand{\hy}{\hat{y}}
\newcommand{\hz}{\hat{z}}
\newcommand{\halpha}{\hat{\alpha}}

\newcommand{\ri}{{\rm i}}

\newcommand{\PrRu}{PrRu$_4$Sb$_{12}$}
\newcommand{\La}{LaOs$_4$Sb$_{12}$}
\newcommand{\Uran}{UPd$_2$Al$_3$}

\newcommand{\MARU}[1]{{\ooalign{\hfil#1\/\hfil\crcr\raise.167ex\hbox{\mathhexbox20D}}}}


The skutterudite \Pra~is a recently found new heavy fermion superconductor.
\cite{Bauer}
It is reported that the time reversal symmetry is broken in the superconducting state,
\cite{Aoki-muSR}
and that there are multiple phases depending on a magnetic field.
\cite{Izawa}
These results together with the absence of the Hebel-Slichter peak
in nuclear quadrupole resonance (NQR) measurements
\cite{Kotegawa}
indicate that unconventional superconductivity is realized in \Pra.
In theoretical studies, an attempt to confirm these experimental results
of unconventional superconductivity has been carried out.
\cite{Miyake,Ichioka,Sergienko,Hotta}

The Pr$^{3+}$ ion in \Pra~has a 4$f^2$ configuration in a $T_h$
(without fourfold axes in a cubic crystal) point-group crystal field.
\cite{Takegahara-Th}
The $\Gamma_1$ singlet ground state is realized
with low-lying $\Gamma_4^{(2)}$ triplet excited states.
This crystal-field configuration can explain naturally
the magnetic field-induced antiferro (AF)-type order above 4.5 T,
\cite{Aoki-field,Ho,Kohgi,Goremychkin}
where the superconductivity disappears.
This is the most important feature compared to the reference superconductors \La~and \PrRu.
\cite{Takeda,Yogi,Frederick}
Thus, we believe that the existence of the low-lying triplet state
is important for superconductivity.

A recent neutron scattering experiment revealed
that the low-lying excitation mode has a weak dispersion relation (exciton).
\cite{Kuwahara-1,Kuwahara-2}
In our previous study, we investigated exciton-mediated superconductivity
by taking only an effective magnetic exchange interaction
between the Pr triplet states and conduction electrons.
\cite{Matsumoto}
In this case, we found an attractive interaction for $d$-wave superconductivity.

The observed exciton dispersion indicates
that there are intersite multipole interactions between the Pr ions.
In the field-induced ordered phase, the following experimental results are reported:
(1) Only a small dipole moment is observed.
\cite{Kohgi}
(2) The critical field exhibits a marked anisotropy depending on the field direction.
\cite{Tayama}
(3) An inelastic neutron scattering experiment
revealed that the intensity of the exciton decreases towards the zone boundary.
\cite{Kuwahara-2}
The above results support a nonmagnetic multipole (quadrupole) interaction.
\cite{Shiina-1,Shiina-2,Shiina-3}
To derive such a substantial multipole interaction via a conduction electron system,
we have to take the orbital degrees of freedom into account
in the exchange interaction between the Pr $4f^2$ and conduction electrons.
\cite{Koga}
Their coupling should contribute to the mass enhancement of the conduction electrons.
\cite{Bauer,Aoki-field,Sugawara}

By neutron scattering measurements,
evidence of an AF multipole intersite interaction was also found.
\cite{Kuwahara-1,Kuwahara-2}
It is known that AF magnetic fluctuations
tend to stabilize singlet pairing superconducting states.
If triplet superconductivity is stabilized,
orbital degrees of freedom are important,
as discussed in the case of the $p$-wave superconductivity of Sr$_2$RuO$_4$.
\cite{Takimoto}
We apply this idea to $f$-electron systems in a different manner
and pursue a stability condition for the triplet superconductivity.
To determine the role of orbitals, \Pra~is a good example,
since its orbital is well defined by the localized nature of the $f$-electron.
In fact, recent NMR and $\mu$SR measurements
reported evidence of odd parity superconductivity.
\cite{Tou,Higemoto}

As mentioned above, the ground state in Pr $4f^2$ is $|\Gamma_1\rangle$
accompanied by low-lying triplet excitations $|\Gamma_{4n}^{(2)}\rangle$ $(n=1,2,3)$.
In the $T_h$ symmetry, the $|\Gamma_{4n}^{(2)}\rangle$ states are
linear combinations of $\Gamma_5$ and $\Gamma_4$ wavefunctions in $O_h$.
\cite{Takegahara-Th,Shiina-1}
\begin{align}
|\Gamma_{4n}^{(2)}\rangle = \sqrt{1-d^2} |\Gamma_5^n\rangle + d |\Gamma_4^n\rangle
\end{align}
Here, n=1,2,3,
and $d$ is the $T_h$ crystal-field parameter
describing the deviation from the $O_h$ symmetry ($d=0$ or $|d|=1$).
The Pr ion is at the center in the cage of Sb$_{12}$.
The $4f^2$ state hybridizes with conduction electrons
having local $a_u$ and $t_u$ symmetries,
which can be formed from the molecular orbitals of Sb$_{12}$.
\cite{Harima}
In this paper, we treat a single $a_u$ dominant conduction band
with strong hybridization with the Pr $f^2$ state.
The $a_u$ electrons hybridize with the $f$ $\Gamma_7$ states of Pr directly
and also with the $f$ $\Gamma_8$ states
via nonlocal mixing with the $t_u$ molecular orbital of Sb$_{12}$.

Let us begin with the following nonlocal exchange interaction
between the 4$f^2$ state of the $i$th Pr and the conduction electrons:
\cite{Koga}
\begin{align}
&H_{\rm ex} = \frac{1}{2} \frac{1}{N} J \sum_{\bsk\bsk'} e^{\ri(\bsk-\bsk')\cdot\br_i}
\label{eqn:H-sd} \\
&\times
\sum_{\lambda\mu}
[ S_{i\lambda} s_\lambda^\mu(\bk,\bk') + Q_{i\lambda} q_\lambda^\mu(\bk,\bk') ]
\sum_{\sigma\sigma'}
\sigma^{\mu}_{\sigma\sigma'} c_{\bsk\sigma}^\dagger c_{\bsk'\sigma'}.
\nonumber
\end{align}
Here, $J=\sqrt{1+20d^2}J_0$ with $J_0$ as the coupling constant for $d=0$.
$S_{i\lambda}$ and $Q_{i\lambda}$ are local magnetic and nonmagnetic operators, respectively,
which couple $|\Gamma_1\rangle$ with $|\Gamma_4^{(2)}\rangle$.
The $\lambda=z, +, -$ components are defined by
\begin{align}
S_z &= (|\Gamma_{42}^{(2)} \rangle \langle \Gamma_1|)
     + (|\Gamma_1 \rangle \langle |\Gamma_{42}^{(2)}|), \cr
Q_z &= (|\Gamma_{42}^{(2)} \rangle \langle \Gamma_1|)
     - (|\Gamma_1 \rangle \langle \Gamma_{42}^{(2)}|), \cr
S_- &= S_+^\dagger = - \sqrt{2} (|\Gamma_1 \rangle \langle \Gamma_{41}^{(2)}|)
                     + \sqrt{2} (|\Gamma_{43}^{(2)} \rangle \langle \Gamma_1|), \cr
Q_- &= Q_+^\dagger = -\sqrt{2} (|\Gamma_1 \rangle \langle \Gamma_{41}^{(2)}|)
                     -\sqrt{2} (|\Gamma_{43}^{(2)} \rangle \langle \Gamma_1|).
\label{eqn:S-and-Q}
\end{align}
In eq. (\ref{eqn:H-sd}), $\mu$ takes $\mu=0,z,+,-$,
where $\sigma^0$ is a unit matrix and $\sigma^\pm = (\sigma^x \pm \ri \sigma^y)/2$.
The momentum dependences of the $\mu$ components of the magnetic
and nonmagnetic channels are described by
$s_\lambda^\mu(\bk,\bk')$ and $q_\lambda^\mu(\bk,\bk')$, respectively.
The former is defined by
\begin{align}
s_z^0(\bk,\bk') &= \ri \frac{3}{7} A \frac{\alpha}{\sqrt{3}} ( A_{\bsk 2} - A_{\bsk' 2} ),~~~
s_z^z(\bk,\bk') = 0, \cr
s_z^+(\bk,\bk') &= \ri \frac{3}{7} \frac{A}{2}
    [ \frac{\alpha}{\sqrt{3}} ( A_{\bsk 1}^* + A_{\bsk' 1}^* )
    + \beta ( A_{\bsk 1} + A_{\bsk' 1} ) ], \cr
s_z^-(\bk,\bk') &= [s_z^+(\bk',\bk)]^*, \cr
s_-^0(\bk,\bk') &= \ri \frac{3}{7} A \frac{\alpha}{\sqrt{3}} ( A_{\bsk 1}^* - A_{\bsk' 1}^* ), \cr
s_-^z(\bk,\bk') &= \ri \frac{3}{7} \frac{A}{2}
    [ - \frac{\alpha}{\sqrt{3}} ( A_{\bsk 1}^* + A_{\bsk' 1}^* )
      + \beta ( A_{\bsk 1} + A_{\bsk' 1} ) ], \cr
s_-^+(\bk,\bk') &= \ri \frac{3}{7} A \beta ( A_{\bsk 2} + A_{\bsk' 2} ), \cr
s_-^-(\bk,\bk') &= \ri \frac{3}{7} A \frac{\alpha}{\sqrt{3}} ( A_{\bsk 2} + A_{\bsk' 2} ).
\label{eqn:sq}
\end{align}
Here, $\alpha$ and $\beta$ are functions of the $T_h$ parameter $d$:
\begin{align}
\alpha=\frac{\sqrt{1-d^2}}{\sqrt{1+20 d^2}},~~~
\beta=\frac{\sqrt{21}d}{\sqrt{1+20 d^2}}.
\end{align}
We introduced the parameter $A$ representing the amplitude of the nonlocal mixing
between the $a_u$ (xyz) and $t_u$ (x,y,z) components of conduction electrons.
The symmetry of the mixing is expressed by the functions $A_{\bsk 1}$ and $A_{\bsk 2}$:
$A_{\bsk 1} = F_{\bsk yz} + \ri F_{\bsk zx}$,
$A_{\bsk 2} = F_{\bsk xy}$,
where
$F_{\bsk xy} = \sin(k_x/2) \sin(k_y/2) \cos(k_z/2)$,
$F_{\bsk yz} = \cos(k_x/2) \sin(k_y/2) \sin(k_z/2)$,
$F_{\bsk zx} = \sin(k_x/2) \cos(k_y/2) \sin(k_z/2)$.
They reflect the symmetry of the bcc lattice.
$s_+^\mu(\bk,\bk')$ is obtained simply by $s_+^\mu(\bk,\bk')=[s_-^{-\mu}(\bk',\bk)]^*$.
Here, $\mu=0,z,+,-$ and $-\mu=0,z,-,+$.
For nonmagnetic channels,
$q_\lambda^\mu(\bk,\bk')$ is obtained from $s_\lambda^\mu(\bk,\bk')$
by replacing $(A_{\bsk' 1}, A_{\bsk' 2}) \rightarrow (-A_{\bsk' 1}, -A_{\bsk' 2})$
for $\lambda = z,-$,
and $(A_{\bsk 1}, A_{\bsk 2}) \rightarrow (-A_{\bsk 1}, -A_{\bsk 2})$ for $\lambda = +$.

Within the above exchange interaction (\ref{eqn:H-sd}),
we obtained the following simple form for a multipole-multipole interaction
for the nearest-neighbor Pr pairs:
\cite{Koga}
\begin{align}
H_{\rm I} = \sum_{\langle ij \rangle} ( D_s \bS_i \cdot \bS_j + D_q \bQ_i \cdot \bQ_j ),
\label{eqn:HI}
\end{align}
since $x$, $y$ and $z$ are identical in the $T_h$ symmetry.
Here, $\bQ_i\cdot\bQ_j=-Q_{iz}Q_{jz}+(Q_{i-}Q_{j+}+Q_{i+}Q_{j-})/2$
due to the anti-Hermite nature of $Q_z$ defined by eq. (\ref{eqn:S-and-Q}).
$\sum_{\langle ij \rangle}$ denotes the summation
over the nearest-neighbor Pr sites on the bcc lattice.
$D_s$ and $D_q$ in eq. (\ref{eqn:HI}) are coupling constants
for the magnetic and nonmagnetic interactions, respectively.
From eq. (\ref{eqn:HI}), we derived the following Hamiltonian for the crystal-field excitons:
\cite{Shiina-3,Matsumoto}
\begin{align}
H_{4f} = \sum_\bsk E_\bsk ( b_{\bsk x}^\dagger b_{\bsk x}
                     + b_{\bsk y}^\dagger b_{\bsk y}
                     + b_{\bsk z}^\dagger b_{\bsk z} ),
\end{align}
where $b_{\bsk \alpha}$ is the bosonic operator for the $\alpha=x,y,z$ component exciton.
$E_\bsk$ is the dispersion for exciton
$E_{\bsk} = \sqrt{ [ \Delta + (D_s + D_q) \Lambda_\bsk ]^2
                 - [(D_s - D_q) \Lambda_\bsk]^2}$,
with $\Lambda_\bsk=8\cos(k_x/2)\cos(k_y/2)\cos(k_z/2)$ for the bcc lattice.
$\Delta$ ($\Delta \gg D_s, D_q$)
is the crystal field splitting from $|\Gamma_1\rangle$ to $|\Gamma_4^{(2)}\rangle$.

For Pr-based systems, it was pointed out
that the exciton plays an important role in mass enhancement
\cite{Fulde}
and superconductivity.
\cite{Ishii}
In this paper, we derive the effective interaction $H_{\rm eff}$
between conduction electrons mediated by the exciton
within the following second-order perturbation:
$H_{\rm eff} = H_{\rm ex} [1/(E_0 - H_0)] H_{\rm ex}$.
Here, $H_0$ is the unperturbed Hamiltonian for the Pr 4$f^2$ and conduction electron systems.
We can eliminate bosonic operators by taking expectation values at low temperatures.
Since we are interested mainly in superconductivity,
we only consider the $c_{-\bsk'}^\dagger c_{\bsk'}^\dagger c_{\bsk}c_{-\bsk}$ type of interaction.
We define the singlet and triplet pair operators by
\begin{align}
\hat{s}_\bsk &= \frac{1}{\sqrt{2}} \sum_{\sigma\sigma'}
        i \sigma_{\sigma\sigma'}^y c_{\bsk\sigma} c_{-\bsk\sigma'}, \cr
\hat{t}_\bsk^\alpha &= \frac{1}{\sqrt{2}} \sum_{\sigma\sigma'}
      ( i \sigma^y \sigma^\alpha )_{\sigma\sigma'} c_{\bsk\sigma} c_{-\bsk\sigma'},
\end{align}
where, $\alpha=x,y,z$.

For singlet pairings, we obtain the effective Hamiltonian
\begin{align}
&H_{\rm eff}^{\rm singlet}
= (\frac{3 J A}{7})^2 ( \frac{1}{3} \alpha^2 - \beta^2 ) \frac{1}{N} \sum_{k k'}
   ( - \frac{U_{s1}}{\Delta} + U_{s2} \frac{\Lambda_{\bsk-\bsk'}}{\Delta^2} ) \cr
&\times
   ( F_{\bsk xy}F_{\bsk'xy} + F_{\bsk yz}F_{\bsk'yz} + F_{\bsk zx}F_{\bsk' zx} )
     \hat{s}_\bsk^\dagger \hat{s}_{\bsk'},
\label{eqn:singlet}
\end{align}
where $U_{s1}=10/9$ and $U_{s2}=2(D_s+D_q/9)$.
For $\alpha^2/3 - \beta^2 > 0$ ($|d|<1/8$),
$d_{xy}$, $d_{yz}$ and $d_{zx}$-waves are stabilized in the lowest order of $1/\Delta$.
For a larger $|d|$, however, the lowest order term becomes repulsive.
For $|d|>1/8$,
we have to consider the second-order term proportional to $\Lambda_{\bsk-\bsk'}/\Delta^2$.
The even parity component of $\Lambda_{\bsk-\bsk'}$ also leads to $d$-wave states.
In this case, the pair wavefunction, for instance, the $d_{xy}$-wave, has a form of
$[\sin(k_x/2)\sin(k_y/2)\cos(k_z/2)]
[\cos(k_x/2)\cos(k_y/2)\cos(k_z/2)]$.
For a small Fermi wave number $k_{\rm F}$,
the $d$-wave is favorable due to large cosine function values.
However, we note that the coupling for the singlet pairing is strongly suppressed
around $|d| \sim 1/8$ ($\alpha^2/3=\beta^2$).

For triplet pairings, an effective interaction similar to eq. (\ref{eqn:singlet}) is derived.
There is no first-order term in $1/\Delta$ for triplet pairings
due to the even parity nature of the functions $(F_{\bsk xy}, F_{\bsk yz}, F_{\bsk zx})$.
The interaction is of the order of $1/\Delta^2$,
and proportional to products of
$(F_{\bsk xy}, F_{\bsk yz}, F_{\bsk zx})$ functions
and the exciton dispersion $\Lambda_{\bsk-\bsk'}$.
There is, for instance, the following term:
\begin{align}
( & F_{\bsk yz} F_{\bsk' yz} \hat{t}_\bsk^{x \dagger} \hat{t}_{\bsk'}^x
+ F_{\bsk zx} F_{\bsk' zx} \hat{t}_\bsk^{y \dagger} \hat{t}_{\bsk'}^y \cr
&
+ F_{\bsk xy} F_{\bsk' xy} \hat{t}_\bsk^{z \dagger} \hat{t}_{\bsk'}^z )
\times \frac{\Lambda_{\bsk-\bsk'}}{\Delta^2}.
\nonumber
\end{align}
Since $\Lambda_{\bsk-\bsk'}$ contains $p$-wave-type terms
such as $\sin(k_x/2)\cos(k_y/2)\cos(k_z/2)$,
$f$-wave pairing functions are obtained
by the product of this $p$-wave-type function and the $d$-wave-type functions
($F_{\bsk xy}$, $F_{\bsk yz}$, $F_{\bsk zx}$).
For a larger $k_{\rm F}$,
the $f$-wave pairing states take advantage against the $d$-wave,
since they contain three sine functions with larger values.

Next, we focus on such triplet pairing states.
We express the effective interaction for triplet pairings in $O_h$ bases as follows:
\begin{align}
H_{\rm eff}^{\rm triplet}
  = \frac{V_{\rm eff}}{\Delta^2} \frac{1}{N} \sum_{kk'} \sum_{\Gamma\Gamma'}
  U_{\Gamma\Gamma'} \hat{g}_\Gamma^\dagger(\bk) \hat{g}_{\Gamma'}(\bk').
\label{eqn:H-eff}
\end{align}
Here, $V_{\rm eff}= (12JA/7)^2 ( D_s + D_q/9 )$
is an effective coupling for triplet pairings.
Both magnetic ($D_s$) and nonmagnetic ($D_q$) multipole-multipole interactions
contribute to the triplet superconductivity.
\begin{align}
\hat{g}_\Gamma(\bk) = V_\Gamma^x(\bk) \hat{t}_\bsk^x
               + V_\Gamma^y(\bk) \hat{t}_\bsk^y
               + V_\Gamma^z(\bk) \hat{t}_\bsk^z
\label{eqn:V}
\end{align}
is an operator for a triplet pair with the $O_h$ representation $\Gamma$.
$\hat{t}^\alpha$ and $V_\Gamma^\alpha(\bk)$ represent
the spin and orbital components, respectively.
$U_{\Gamma\Gamma'}$ represents an interaction connecting $\Gamma$ and $\Gamma'$.
We introduce a $\bd$-vector by a linear combination of
$\bd(\bk)= \sum_\Gamma \Delta_\Gamma \bV_\Gamma(\bk)$.
Here, $\Delta_\Gamma$ is the amplitude of the order parameter for $\Gamma$.
The order parameter $\Delta_\Gamma$ can be obtained by solving the following gap equation:
\begin{equation}
\Delta_\Gamma = - 2\pi T N_0 \sum_m \frac{1}{|\omega_m|}
  \sum_{\Gamma'\Gamma''} U_{\Gamma\Gamma'}
  \langle \bV_{\Gamma'}(\bk) \cdot \bV_{\Gamma''}(\bk) \rangle_k \Delta_{\Gamma''}.
\label{eqn:gap-eq}  
\end{equation}
Here, $\langle \cdots \rangle_\bsk$ represents an integral over the Fermi surface.
$\omega_m$ is a Matsubara frequency for fermions.
$N_0$ is the density of states at the Fermi energy.
In the $T_h$ system, the gap equations can be reduced to
the $\Gamma_1\oplus\Gamma_2$, $\Gamma_{31}\oplus\Gamma_{32}$ and $\Gamma_4\oplus\Gamma_5$ types.
We find that the transition temperatures for these pairing states are similar.
We focus on one of the pairing types ($\Gamma_4\oplus\Gamma_5$) in this paper,
since this three-dimensional representation
is most appropriate for the observed experimental results.
The $V_\Gamma^\alpha(\bk)$ functions in eq. (\ref{eqn:V})
for the $\Gamma_4\oplus\Gamma_5$ type are listed in Table I.
The interaction $U_{\Gamma\Gamma'}$ in eq. (\ref{eqn:gap-eq}) is given in Table II.
The off-diagonal matrix elements are proportional to $\alpha\beta$.
They vanish in the $O_h$ system.
The eigenvalues of the matrix in eq. (\ref{eqn:gap-eq})
determine the effective interaction corresponding to the obtained eigenmode.
A negative (positive) eigenvalue means an attractive (repulsive) interaction.
The lowest eigenvalue leads to the highest transition temperature.
The eigenvalue depends on the $T_h$ parameter $d$ strongly via the off-diagonal components.

\begin{table}[t]
\caption{
Orbital component of the triplet pair operator.
Here, $x$, $y$ and $z$ represent
$x = \sin(k_x/2)$, $y = \sin(k_y/2)$ and $z = \sin(k_z/2)$, respectively.
In the same manner,
$x' = \cos(k_x/2)$, $y' = \cos(k_y/2)$ and $z' = \cos(k_z/2)$.
Since the $\Gamma_4\oplus\Gamma_5$ type has a three-dimensional representation,
there are other two components
obtained by replacing $(x,y,z)\rightarrow(y,z,x)$ cyclicly.
There are two basis functions (denoted by A and B) for the $\Gamma_4$ type
and three basis functions (denoted by A, B and C) for the $\Gamma_5$ type.
}
\begin{displaymath}
\begin{array}{|c|c|c|c|} \hline
    & V_\Gamma^x(\bk) & V_\Gamma^y(\bk) & V_\Gamma^z(\bk) \\ \hline
\hat{g}_{\Gamma_x^{\rm A}} & 0 & \frac{1}{\sqrt{2}} x^2  y'^2 z z' &
                                -\frac{1}{\sqrt{2}} x^2 y y' z'^2  \\
\hat{g}_{\Gamma_x^{\rm B}} & 0 & - \frac{1}{2}  x'^2 y^2 z z' &  \frac{1}{2}  x'^2 y y' z^2 \\
\hat{g}_{\Gamma_{yz}^{\rm A}} & - \sqrt{\frac{2}{3}} x x' y y' z z' &
           \frac{1}{\sqrt{6}} x^2  y'^2 z z' &
           \frac{1}{\sqrt{6}} x^2 y y' z'^2 \\
\hat{g}_{\Gamma_{yz}^{\rm B}} & x x' y y' z z' &
         \frac{1}{2}  x'^2 y^2 z z' & \frac{1}{2}  x'^2 y y' z^2 \\
\hat{g}_{\Gamma_{yz}^{\rm C}} & x x' y y' z z' &
       - \frac{1}{2}  x'^2 y^2 z z' &
       - \frac{1}{2}  x'^2 y y' z^2 \\ \hline
\end{array}
\end{displaymath}
\end{table}

\begin{table}[t]
\caption{
Matrix elements of $U_{\Gamma\Gamma'}$.
$a_1=(3/2)\alpha^2-(1/2)\beta^2$,
$a_2=-(1/6)\alpha^2+(1/2)\beta^2$,
$a_3=(2/\sqrt{3})\alpha\beta$,
$a_4=3/2$ and
$a_5=(4/3)\alpha^2$.
}
\begin{displaymath}
\begin{array}{|c|ccccc|} \hline
    & \Gamma_x^{\rm A} & \Gamma_x^{\rm B} & \Gamma_{yz}^{\rm A} & \Gamma_{yz}^{\rm B} & \Gamma_{yz}^{\rm C} \\ \hline
\Gamma_x^{\rm A}    & a_4 & 0  & 0 & 0 & 0 \\
\Gamma_x^{\rm B}    & 0 & a_5  & 0 & -a_3 & a_3 \\
\Gamma_{yz}^{\rm A} & 0 & 0  & a_4 & 0 & 0 \\
\Gamma_{yz}^{\rm B} & 0 & -a_3  & 0 & a_1 & 0 \\
\Gamma_{yz}^{\rm C} & 0 & a_3 & 0 & 0 & a_2 \\ \hline
\end{array}
\end{displaymath}
\end{table}

Let us discuss effects of the $T_h$ symmetry on superconductivity.
As shown in Table II, the off-diagonal terms connect the lowest eigenstate
with higher-lying eigenvalue states.
This effect is similar to the level repulsion between the states.
As a result, the lowest eigenvalue is reduced compared with
that in the absence of the off-diagonal terms,
and the effective attractive interaction is enhanced as in Fig. 1.
Therefore, the $T_h$ symmetry stabilizes the triplet superconductivity.
In contrast to the triplet pairing states,
there is no such effect for the $d$-wave states [see eq. (\ref{eqn:singlet})].

Next, we apply our result to explain the recent experimental data.
Zero-field $\mu$SR measurements revealed an internal magnetic field
emerging at the superconducting transition temperature.
\cite{Aoki-muSR}
This indicates that superconductivity
with a broken time reversal symmetry is realized in \Pra,
implying the multicomponent $\Gamma_{31}\oplus\Gamma_{32}$ or $\Gamma_4\oplus\Gamma_5$ type.
For the $\Gamma_4\oplus\Gamma_5$ type,
there are threefold degenerate solutions.
There are two types of combination with a broken time reversal symmetry.
One is the $\bd_{\Gamma_{4y}}(\bk)\pm\ri\bd_{\Gamma_{4z}}(\bk)$ type,
and the other is the
$\bd_{\Gamma_{4x}}(\bk)+\omega\bd_{\Gamma_{4y}}(\bk)+\omega^2\bd_{\Gamma_{4z}}(\bk)$ type
with $\omega=e^{\pm \ri 2\pi/3}$.
These states with a broken time reversal symmetry are nonunitary states
with a finite $\bq(\bk)=\ri\bd(\bk)^*\times\bd(\bk)$ vector.

Let us discuss the $T_h$ nature of the gap function $\sqrt{|\bd(\bk)|^2-|\bq(\bk)|}$
for the nonunitary states.
In Fig. 2(a), we show the gap function,
for instance, $\bd(\bk)=\bd_{\Gamma_{4z}}(\bk)$
for a $T_h$ parameter $\alpha=\beta=1/\sqrt{2}$ ($d=1/\sqrt{22}$).
Although the shape looks like a clover with four leaves,
there are six point nodes along all the basal axes $x$, $y$ and $z$.
The fourfold symmetry around the $z$ axis is broken under the $T_h$ symmetry.
The angle of one leaflike shape tends to be small towards the $y$ direction,
and the gap decreases strongly in the $x$ direction.
The other two states [$\bd_{\Gamma_{4x}}(\bk)$ and $\bd_{\Gamma_{4y}}(\bk)$] can appear
with a broken time reversal symmetry,
since the three states are degenerate.
As shown in Fig. 2(b),
the $\bd_{\Gamma_{4y}}$ component is combined with $\bd_{\Gamma_{4z}}$,
and the gap function takes a value close to the $y$ direction,
which is larger than that for the single component of $\bd_{\Gamma_{4z}}$.

Multiple superconducting phases are reported by thermal conductivity measurements.
\cite{Izawa}
In the lower-field phase,
it was reported that the gap function has a twofold symmetry around the $z$ axis,
\cite{Izawa,Huxley}
which may correspond to the nonunitary triplet pairing states
$\bd(\bk)=\bd_{\Gamma_{4y}}(\bk)\pm\ri\bd_{\Gamma_{4z}}(\bk)$ discussed above.

\begin{figure}[t]
\begin{center}
\includegraphics[width=8cm]{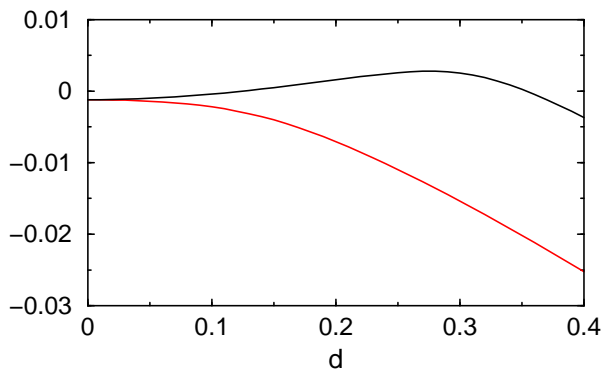}
\end{center}
\caption{
$T_h$ parameter $d$ dependence of the effective interaction
for $\Gamma_4 \oplus \Gamma_5$ type in arbitrary units.
$k_{\rm F}=0.75\pi$ is considered here.
Red and black lines are the eigenvalues
with and without the off-diagonal interaction terms, respectively.
}
\end{figure}

\begin{figure}[t]
\begin{center}
\includegraphics[width=4cm]{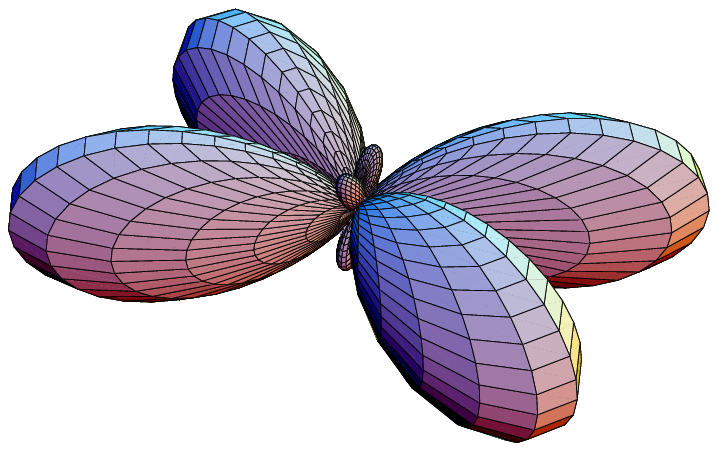}
\includegraphics[width=4cm]{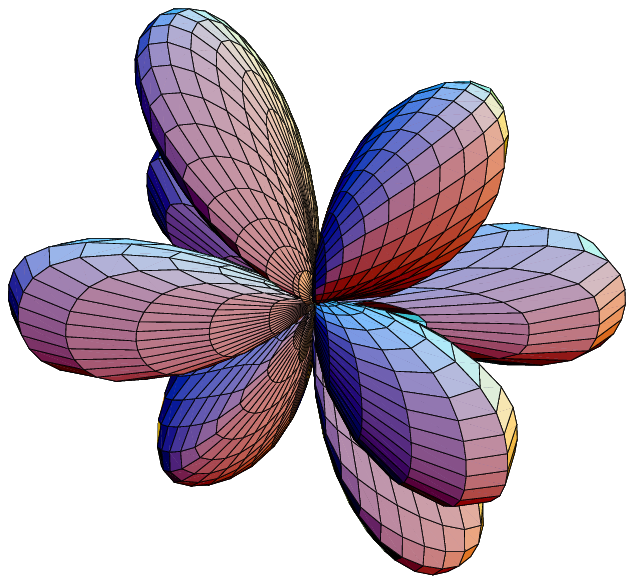}
\end{center}
\caption{
Gap functions at $\alpha^2=\beta^2=1/2$ ($d=1/\sqrt{22}$) for $k_{\rm F}=0.75\pi$:
(a) $\bd(\bk)=\bd_{\Gamma_{4z}}(\bk)$ and
(b) $\bd(\bk)=\bd_{\Gamma_{4y}}(\bk)\pm\ri\bd_{\Gamma_{4z}}(\bk)$.
}
\label{fig:gap-4-5}
\end{figure}

In our previous study, we found the order of $J_0^2 D^2 / \Delta^3$
for the coupling of the $d$-wave pairing interaction,
where only an effective magnetic exchange interaction
between the triplet excitations and conduction electrons was considered.
\cite{Matsumoto}
Here, $D$ ($\ll\Delta)$ represents the width of the exciton dispersion.
In contrast to this, we considered both the $a_u$ and $t_u$ local symmetries
in the single conduction band,
where the latter has orbital degrees of freedom.
This model leads to attractive channels for triplet pairing
with the order of $A^2 J_0^2 D / \Delta^2$.

If the field-induced order is a quadrupolar type, strong orbital fluctuations are expected.
Thus, the parameter $A$ can have a sufficiently large value of $A > \sqrt{D/\Delta}$.
In this case, the superconducting state discussed in this paper can be stabilized.
The triplet pairing is promising when the singlet pairing interaction is suppressed
[$|d| \sim 1/8$ in eq. (\ref{eqn:singlet})].
Since the spin-orbit interaction works in the exchange process via the $4f$ orbitals,
the spin of the conduction electron is not conserved in the scattering with orbital exchange.
This can give rise to attractive triplet channels
even when the exciton has AF fluctuations.
We also found that the $T_h$ symmetry stabilizes the triplet superconductivity
and reflects the symmetry of the gap function.
It is remarkable that the atomic nature of Pr is revealed by the superconductivity.

We would like to express our sincere thanks to H. Shiba for valuable discussions
and critically reading the manuscript.
We thank Y. Aoki, W. Higemoto, K. Izawa, M. Kohgi, H. Kusunose, K. Kuwahara, O. Sakai, R. Shiina,
T. Takimoto and H. Tou for useful discussions.
We acknowledge M. Oichi for his kind assistance.
This work is supported by JSPS Grants-in-Aid for Scientific Research in Priority Area
`Skutterudite' (No. 16037207) and for Young Scientists (No. 16740197).


\end{document}